# Anisotropy-axis orientation effect on the magnetization of $\gamma$-$Fe_2O_3$ frozen ferrofluid

S. Nakamae[1], C. Crauste-Thibierge[1], K. Komatsu[1], D. L'Hôte[1], E. Vincent[1], E. Dubois[2], V. Dupuis[2] and R. Perzynski[2]

[1]Service de Physique de l'Etat Condensé (CNRS URA 2464) DSM/IRAMIS, CEA Saclay F-91191 Gif sur Yvette, France
[2]Physicochimie des Electrolytes, Colloïdes et Sciences Analytiques, UMR 7195, Université Pierre et Marie Curie, 4 Place Jussieu, 75252 Paris, France

E-mail: sawako.nakamae@cea.fr

**Abstract.** The effect of magnetic anisotropy-axis alignment on the superparamagnetic (SPM) and superspin glass (SSG) states in a frozen ferrofluid has been investigated. The ferrofluid studied here consists of maghemite nanoparticles ($\gamma$-$Fe_2O_3$, mean diameter = 8.6 nm) dispersed in glycerine at a volume fraction of ~15%. In the high temperature SPM state, the magnetization of aligned ferrofluid increased by a factor varying between 2 and 4 with respect to that in the randomly oriented state. The negative interaction energy obtained from the Curie-Weiss fit to the high temperature susceptibility in the SPM states as well as the SSG phase onset temperature determined from the linear magnetization curves were found to be rather insensitive to the anisotropy axis alignment. The low temperature aging behaviour, explored via 'zero-field cooled magnetization' (ZFCM) relaxation measurements, however, show distinct difference in the aging dynamics in the anisotropy-axis aligned and randomly oriented SSG states.

## 1. Introduction

Ferrofluids are composed of nanometre scale ferro- or ferrimagnetic particles such as maghemite and magnetite that are suspended in a fluid carrier. When diluted, these particles are small enough (diameter typically below 10 nm) to be dispersed uniformly within a carrier fluid and their thermal fluctuations contribute to the bulk superparamagnetic response of the frozen fluid at high enough temperatures. Soon after the discovery of ferrofluids, it was recognized that the inter-particle dipole-dipole interactions and the polydispersity of nanoparticle sizes lead to equilibrium magnetization curves which cannot be approximated by an assembly of individual monodisperse superspins. Furthermore, when sufficiently concentrated, interparticle interactions were found to produce a collective state at low temperatures (usually well below the freezing point of the carrier fluid), showing similarities with atomic spin-glasses [1,2]. Subsequently, experimental results in support of such disordered collective states, called superspin glass (SSG), have been obtained [3-7]. The SSG state is believed to be the product of the random distributions of positions, sizes and anisotropy-axis orientations of magnetic nanoparticles that interact with each other via dipolar interactions. The dipolar field falls off as $r^{-3}$ and therefore, it is of long range nature. Furthermore, the microscopic 'flip-time' of one superspin (in the order of $10^{-9}$ sec) is much longer than an atomic spin flip time (in the order of $10^{-12}$ sec). These features differentiate the physics of SSG phase from that of atomic spin-glass phases. Nevertheless, theoretical models developed for atomic spin-glass have so far succeeded in describing many aspects of SSG dynamics. The slow dynamics of SSG's is of particular interest because a much shorter time scale becomes experimentally accessible with SSG's. An example that can illustrate the advantage of such a long flip-time is the slow growth of a dynamical correlation length in spin glass phases. Numerical simulations on the growth behaviour of correlation length exist [8-10]; however, a direct comparison between the experimental data and these predictions is difficult due to a large gap between the usual time scales explored by numerical simulations and that accessible in laboratory experiments on atomic spin glasses [11,12]. With longer flip times one can hope to bridge the gap between experiments and theories [7].

Another advantage of using concentrated frozen ferrofluids is the easy-access to key physical parameters that strongly influence the SSG phase, such as the interaction energy, the individual superspin size and the anisotropy alignment. In magnetically aligned frozen ferrofluids, not only the positions of all particles are fixed in space but also their magnetic easy-axes are uniformly oriented parallel to the external bias field direction. Therefore, the distribution of anisotropy axes is no longer random. The effect of anisotropy-axis alignment on the physical properties of nanoparticle assemblies have been studied both theoretically and experimentally in their superparamagnetic state [13-19]. However, little is known on its influences at low temperatures in the presence of dipole-dipole interactions (*i.e.*, high concentrations) [20-23]. Due to the loss of a disorder in the anisotropy orientation distribution, the SSG phase of a magnetically aligned frozen ferrofluid may well behave differently from that of randomly oriented nanoparticles.

In this study we have used a ferrofluid consisting of maghemite, $\gamma$-$Fe_2O_3$ nanoparticles dispersed in glycerine and aligned the easy magnetization axis of individual nanoparticles by freezing glycerine in the presence of high magnetic fields ($H > 15$ kOe). After performing a series of magnetization measurements (DC magnetization, ac susceptibility and low temperature magnetization relaxation) the ferrofluid was warmed up to above the melting temperature of glycerine to destroy the anisotropy-axis alignment. Then the same series of experiments were repeated on the same ferrofluid, this time with particle's anisotropy axis distributed randomly. As the anisotropy-axis alignment is the only difference between the two sets of measurements, the direct comparison between the two should elucidate exclusively its influence on their magnetic behaviour in both the SPM and the SSG states.

This paper is organized as follows. Section 2 is devoted to the sample description and the experimental methods used in our study. In section 3, phenomenological models used to analyze our experimental data are discussed. The experimental data analysis and the discussion are given in section 4. A brief summary of our findings is found in the last section.

## 2. $\gamma$-$Fe_2O_3$ ferrofluid sample and experimental methods

*2.1. Ferrofluid Sample and anisotropy axis alignment*

The ferrofluid used in the present study is composed of maghemite, $\gamma$-$Fe_2O_3$, nanoparticles dispersed in glycerine at ~15% volume fraction. The distribution of the nanoparticles' diameters can be described by a log-normal distribution of characteristics; *i.e.*, mean diameter $d_o$ =8.6 nm (ln($d_o$) = <ln($d$)>) and dispersion $\sigma$=0.23 [24]. Due to their small sizes, these nanoparticles are magnetic single-domains with an average permanent magnetic moment of ~$10^4 \mu_B$. Approximately, 1.5 L of ferrofluid was sealed hermetically inside a small glass capillary (1mm inner diameter). The magnetization and the ac susceptibility measurements were performed using a commercial SQUID magnetometer (CRYOGENIC™ S600).

In order to physically rotate and align nanoparticles' anisotropy axes, an external bias field $H$ (15 and 30 kOe) was applied at 300K for over one hour. These values were chosen based on the birefringence measurements conducted on a concentrated ferrofluid similar to ours where an axis-alignment at $H > 5$ kOe at room temperature was observed [25]. The ferrofluid was cooled down to 150K (< 190 K = freezing temperature of glycerine) before removing the strong bias field. DC magnetization was then measured as a function of temperature with a 1 Oe applied field. The magnetization curves obtained from the sample aligned under 15 and 30 kOe were found to superimpose over one another within the experimental uncertainty, indicating that a uni-axial anisotropy orientation is achieved [22]. All data presented on the 'aligned' sample hereafter were taken on the ferrofluid aligned at 30 kOe. The frozen ferrofluid with randomly oriented nanoparticles is referred to as 'random' sample.

The comparison of 'aligned' and 'random' samples implies to have the knowledge of the microscopic structure of the samples, especially under magnetic fields. This has been widely explored in several previous studies [25-27]. Coupled Small Angle Scattering and magneto-optical measurements [25] proved that the properties of the magnetic nanoparticle dispersions are controlled by several parameters; the dipolar parameter $\gamma/\Phi$, the osmotic pressure $\Pi$, and the volume fraction $\Phi$. These parameters define the location of the sample in the dispersion phase diagram, which mainly depends on the interparticle interactions. In these systems, the van der Waals attractions and the dipolar magnetic interactions (attractive on average) between the nanoparticles are counterbalanced by the electrostatic repulsion created by the surface charges; citrate molecules adsorbed on the nanoparticle surface. The pressure $\Pi$ is essentially controlled by the electrostatic interaction, and the dipolar interaction can be quantified by $\gamma = \dfrac{\mu_o \mu^2}{\bar{r}^3 k_B T}$, ratio between the magnetic dipolar energy and the thermal energy, $k_B T$ ($\mu$: dipole moment of the particle, $r$: mean distance between particles). For the sample used here with $\Phi \sim 15\%$ and the salt concentration is 0.05 M, $\gamma/\Phi$ equals 20 at 300K and this value grows to 32 at 190K and to 40 at 150K. In glycerine as well as in water, no aggregates are formed under such conditions in similar nanoparticle dispersions, even in the presence of a strong magnetic field [25-27]. Therefore the ferrofluid studied here is most likely to be an aggregation-free dispersion of individual particles even under a strong magnetic field and at low temperature. Note that under a strong field, the structure nevertheless becomes slightly anisotropic because the interparticle interactions become anisotropic due to the orientation of the magnetic dipoles. However, the mean distance between the nanoparticles is found to remain isotropic within the resolution of neutron scattering [26].

*2.2. Magnetization Measurements*

In order to understand the effect of anisotropy axis alignment on the high temperature SPM phase as well as on the low temperature SSG aging dynamics of a frozen ferroluid, we have carried out a series of measurements including: low field DC magnetization (zero-field cooled (ZFC) and field cooled (FC)) vs. temperature, ac magnetic susceptibility vs. temperature (with an excitation field of 1Oe oscillated at frequencies between 0.04 and 8Hz) and the zero-field cooled magnetization (ZFCM) relaxation at temperatures below $T_g$. The experimental procedure for ZFCM relaxation measurements is as follows. First, the samples are cooled from a temperature (140 K) well above the superspin-glass transition temperature, $T_g \sim 70$ K (for both SSG states) to the measuring temperature, $T_m = 49$K ($\sim 0.7\ T_g$) in zero field. After waiting for a period of $t_w$ (waiting time ranging between 3 and 24 ks), a small probing field (0.15 Oe $\leq H \leq$ 8 Oe) is applied at $t = 0$. The magnetization relaxation toward a final value, $M_{FC}$ (field cooled magnetization) is measured over a long period time, $t$, during which the relaxation rate also evolves, continuously changing the slope of the ZFCM response function. In the case of aligned SSG, measurements at 59.5 K ($\sim 0.84\ T_g$) were also performed.

## 3. Phenomenological models and data analysis methods

*3.1. Magnetization relaxation scaling*

Key physical phenomena of interest here related to the aging in the SSG states are the time dependent magnetization relaxation and the associated relaxation rates. In atomic spin glasses, both the thermoremanent magnetization (TRM) and the zero-field cooled magnetization after a temperature quench in the spin-glass phase can be expressed as a sum of a stationary equilibrium term, $m_{eq}(t)$, and an aging term, $m_{ag}(t, t_w)$.

$$\frac{M}{M_{FC}} = m_{eq}(t) + m_{ag}(t,t_w) = \pm A\left(\frac{\tau_o}{t}\right)^\alpha + f\left(\frac{\lambda}{t_w^\mu}\right) \quad (1),$$

where $A$ is a prefactor which takes a positive value in the case of TRM and a negative value for ZFCM, $\tau_o$ is a microscopic 'spin-flip' time, $\alpha$ and $\mu$ are scaling exponents. $\lambda/t_w^\mu$ with $\lambda = t_w[(1+t/t_w)^{1-\mu}-1]/[1-\mu]$ is an effective time variable which takes in account the $t_w$ dependent evolution of the magnetization relaxation [28,29]. When fitting parameters ($\mu$, $\alpha$ and $A$) are properly chosen, $m_{ag}(t,t_w)$ of spin-glass magnetization; i.e., $M/M_{FC}$ - $m_{eq}$ at different $t_w$'s all collapse onto a single master curve function of $\lambda/t_w^\mu$. Values of $\mu \neq 1$ indicate by how much the 'effective age' of a spin glass deviates from its 'nominal age'; that is, experimental waiting time, $t_w$.

In the magnetization relaxation of superspin glasses made of interacting fine magnetic nanoparticles, an additional non-aging, time-logarithmic term has been identified [6,30]. This relaxation term, $B\log(t/\tau_o)$ is believed to stem from superparamagnetic moments that do not participate in the superspin glass aging dynamics and must be treated independently. The scaling of low temperature ZFCM curves would serve as an additional indication of a SSG phase in frozen ferrofluids with or without the anisotropy-axis alignment.

*3.2. Magnetization relaxation rate, effective age of a (super)spin glass and dynamic spin correlations*

In a spin glass, the magnetization relaxation rate ($S$) after an external field change is often expressed as a log-derivative of $M/M_{FC}$, i.e., $S = d(M/M_{FC})/d\log(t)$. $S(\log(t))$ contains a maximum reached at a characteristic time, $t_w^{eff}$ that corresponds to the time at which the relaxation rate becomes the fastest, $S_{max}$. The quantity $S(\log(t))$ is equivalent to the relaxation time distribution of dynamically correlated (super)spin zones [31], and thus $t_w^{eff}$ is commonly referred to as the *effective age* of the system since the temperature quench time. A wide spread of $S(\log(t))$ is indicative of the slow and non-exponential relaxation of the response function in a (super)spin glass state.

One can extract both qualitative and quantitative information on the dynamics of (super)spin correlations (number and length) in the glassy phase by studying the $t_w^{eff}$ position shift in response to the changes in experimental control parameters, $t_w$ and $H$ via ZFCM measurements. This experimental approach relies on the assumption that the observed reduction in the effective age of the system upon the change in an external magnetic field is due to the Zeeman energy ($E_Z(H)$) coupling to many subsets of dynamically correlated (super)spins [11,12,32]. At $t = t_w$ after a temperature quench in zero field, a typical size of the correlated spins has grown to $N_s(t_w)$ with an associated free energy barrier of $E_B(t_w)$. The relaxing of $N_s(t_w)$ dynamically correlated (super)spins toward their final state requires a cooperative flip of all $N_s(t_w)$. Therefore, in response to a vanishingly small external field, such a cooperative flipping should equally require an amount of time $\sim t_w$:

$$t_w(H \sim 0) = \tau_0 \exp(E_B(t_w)/k_BT) \quad (2),$$

where $\tau_o$ is, once again, a microscopic flipping time of a single (super)spin. Indeed in atomic spin glasses and in one randomly oriented superspin glass, $S_{max}$ occurs at a characteristic time $t \sim t_w$ at very low fields. In the presence of a small but non-negligible $H$, however, $E_Z(H)$ acts to reduce the barrier energy to a new value, $E_B(t_w)-E_Z(t_w, H)$ by coupling to $N_s(t_w)$ correlated spins. Therefore, one expects a shift of the $S_{max}$ position to shorter times $t_w^{eff}(H) < t_w$.;

$$t_w^{eff}(H) = \tau_0 \exp\{(E_B(t_w)-E_Z(H,t_w)/k_BT)\} \quad (3).$$

By combining the expressions (2) and (3), the relationship between the relative decrease in $t_w^{eff}$ (effective age) with respect to $t_w$ (nominal age) of the system and the Zeeman energy exerted onto the $N_s$ correlated (super)spins can be written as,

$$\ln\left(\frac{t_w^{eff}}{t_w}\right) = -\frac{E_Z(t_w, H)}{k_B T}. \qquad (4).$$

$E_Z(H,t_w)$ depends on both the external field and the number of correlated spins, $N_s(t_w)$. Once $E_Z(H)$ is determined, $N_s$ may be extracted knowing that $E_Z(H) = M(N_s)H$. The exact form of $E_Z$ is not readily known and therefore it is often speculated from the experimental observations [11,12]. In the case of Ising-type spin glasses $E_Z(H)$ was found to grow linearly with $H$, while in Heisenberg spin glasses, a quadratic dependence on $H$ was reported. These experimental observations were interpreted to reflect $E_Z(H) = \sqrt{N_s}\mu H$ in Ising spin glasses (with relatively small values of $N_s$, see [11] for more detail) and $E_Z = N_s\chi_{FC}H^2$ in the case of Heisenberg-like spins (with macroscopically large values of $N_s$) where $\mu$ is the magnetic moment of one spin and $\chi_{FC}$ is the field cooled susceptibility per (super)spin. The ZFCM method has been used successfully in atomic spin-glasses [11,12] and lately in a randomly oriented superspin glass by our group [7]. Our previous ZFCM experiments performed on a random SSG system exhibited closer to a quadratic dependence on $H$, and the $N_s$ values were extracted based on Heisenberg spin glass model accordingly.

## 3. Results and Discussion

*3.1. Anisotropy axis alignment effect on the superparamagnetic behaviour*

In figure 1, the Zero-Field Cooled (ZFC)/Field Cooled (FC) DC susceptibility curves (*M/H*) of the frozen ferrofluid with and without anisotropy axis alignment are presented. 1 Oe probing field was used in both measurements. Here, we have taken in account the demagnetization factor ~0.3 due to a short cylindrical shape of our sample [33]. Notice that due to the melting of glycerine starting around 200 K and above, the $\chi(T)$ of the aligned sample approaches that of the random sample. Below 200 K where superspins are physically blocked, the $\chi$ of aligned sample becomes considerably larger than that in the random state.

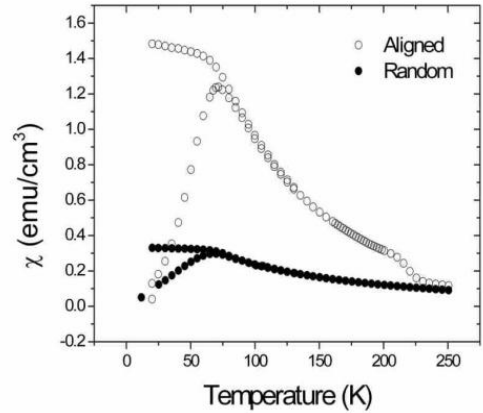

**Figure 1.** ZFC and FC DC magnetic susceptibility curves of $\gamma$-Fe$_2$O$_3$ ferrofluid in aligned and random states. An external field of 1 Oe was used in both measurements. Note that *M(T)* at *T* > 150 K in the aligned sample were taken at the end of all other magnetization measurements presented in this study.

In the case of 'non-interacting' and monodisperse superparamagnetic particles, $M_{//}$, magnetization in the direction of an external field of a randomly aligned ferrofluid at high $T$ follows the Langevin function [34], $M_{//}(\xi)=M_s[\coth(\xi)-1/\xi]$ where $M_s$ is the saturation magnetization of the magnetic material and $\xi = \mu H/k_B T$ ($\mu = V_p M_s$ is the magnetic moment of each particle with $V_p$ being the volume of one nanoparticle). In a weak field, high temperature limit $M_{//}(\xi)$ becomes $N\mu^2 H/3Vk_B T$ (Curie Law). If all particles' magnetic anisotropy axes are oriented parallel to an externally applied field, magnetization is no longer given by the Langevin law. In the extreme limit where anisotropy energy $E_a \to \infty$ and without interactions, $M_{//} = M_s \tanh(\xi)$ which becomes $N\mu^2 H/Vk_B T$ in the weak field limit [35]. The anisotropy energy of our maghemite nanoparticles, $E_a/k_B = 2\times 300K$ [36] is much greater than the magnetic energy $\xi T \sim 1$ K (for $H$ in the order of 1 G).

In the presence of dipole-dipole interactions, each nanoparticle responds to its total local field, $H_T$ which is a sum of applied magnetic field and the dipolar fields exerted by the surrounding superspins near and far. Therefore for the total local field for a nanoparticle located at $x_i$, one has $H_T(x_i) = H_{ext} + H_{diople}(x_i)$. Jönsson and Garcia-Palacios have calculated the linear equilibrium susceptibility $\chi$ in weakly interacting superparamagnets.[4,37] In their work, $\chi$ is expressed in the form of an expansion with coefficients that depend on dipolar interactions as well as on anisotropy effects. The results indicated that (in the absence of an external bias field) all traces of anisotropy are erased in the linear susceptibility of a superparamagnetic system with randomly distributed anisotropy axes and the expression for isotropic spins ($N\mu^2/3Vk_B T$) is recovered. For systems with parallel aligned axes, the dipolar interactions were found to be stronger and the corresponding low temperature susceptibility approaches that of Ising spins; i.e., $N\mu^2/Vk_B T$. As seen in figure 1, the ratio between the $\chi(T)$ of the aligned frozen ferrofluid to that of the randomly oriented ferrofluid is approximately 2 at 200 K and this value grows to about 4 at the ZFC maximum temperature. The ratio between the two susceptibility values in the SPM regime that exceeds 3 may indicate that the dipole-dipole interactions in the present ferrofluid are beyond the weak interaction limit. The interparticle dipolar interactions are known to play an important role in concentrated magnetic nanoparticle systems and can lead to an increase > 3 of the linear susceptibility from the Langevin value [38,39]. Therefore, a change in dipolar interaction energy due to the anisotropy axis alignment may explain the apparent increase in the linear $\chi$ observed here. However, the transition temperature, loosely defined here as the temperature at which the ZFC and FC curves separate, is found at ~70K in both systems. As the $T_g$ is known to depend strongly on the dipolar interactions (i.e., concentrations) the insensibility of $T_g$ to the anisotropy alignment disproves a significant change in dipolar interaction energy speculated above.

To further elucidate the change in the interaction strength, we have plotted $1/\chi$ of the high temperature SPM phase as a function of temperature in order to extract the (negative) interaction energy appearing in the form of the Curie-Weiss law; $\chi(T) \propto (T - T_o)^{-1}$. The value of $T_o$ in the aligned ferrofluid = -15 K ± 10 is not very different from that found in the random state = -25K ± 3. Note that an arbitrary and temperature independent (diamagnetic) contribution needed to be subtracted from the raw data to perform these fits. Additionally, the upper bound of the experimentally accessible SPM temperature range is limited by the melting of glycerine near 200 K. These facts contributed to large uncertainties in $T_o$. It is nevertheless interesting to consider the ratio between the susceptibilities in the aligned and random samples (~3.5 between 200 and 100 K, see figure 2). As a function of $(T - T_o)$ with their respective $T_o$ values (Inset figure 2), the ratio becomes 3.15, approaching the theoretical value of 3. In disordered systems such as ferrofluids studied here, the physical meaning of the negative interaction energy is not easily understood. It has been previously demonstrated by Chantrell et al., [40] that the negative interaction energy (extracted from high temperature SPM simulation on interacting nanoparticle systems) depends strongly on the packing density of fine magnetic particles as well as on the system geometry; i.e., long-range interactions. Therefore, the lack of a discernible change in $T_o$ suggests dipolar interaction remains rather constant under the anisotropy alignment change.

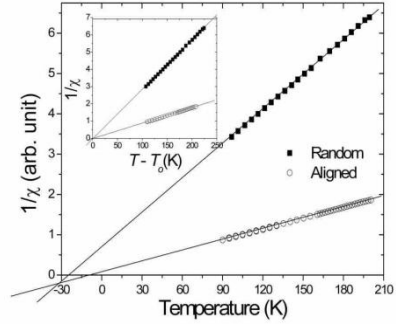

**Figure 2.** $1/\chi$ vs. temperature in the high temperature SPM region. The x-axis intercepts indicate the values of $T_o$. A diamagnetic and temperature independent contribution $M_o$, presumably due to the sample holder (glass capillary) needed to be subtracted from the raw data for this analysis. The inset shows the same $1/\chi$ plotted against $T - T_o$ ($T_o = -15$ and $-25$ K are used for the aligned and the random states, respectively)

*3.2. Persistence of SSG state in an aligned ferrofluid at low temperature*

In order to differentiate the SSG transition from the SPM blocking behaviour, frequency ($\omega$) dependence of AC susceptibility was measured and the peak temperature $T_g(\omega)$ at which the real part of susceptibility reaches its maximum value was analyzed. If the frozen ferrofluid in either form is an ensemble of independent superparamgnetic centres, $T_g(\omega)$ can be fitted to the Arrhenius law: $\omega^{-1} = \tau_o \exp\left(E_a / k_B T_g(\omega)\right)$, with a physically reasonable value of $\tau_o$ (in the order of $10^{-9} \sim 10^{-10}$ s for the types of magnetic particles studied here). The fits to the Arrhenius law give unphysical values of $\tau_o$ $\sim 10^{-19} \sim 10^{-20}$ s in both cases indicating possible phase transitions at taking place at $T_g(\omega)$. A second order phase transition (divergence of a correlation length) toward a disordered state exhibits a critical behaviour [41] that is described by

$$\omega^{-1} = \tau_0^* \left[\frac{T_g(\omega) - T_g}{T_g}\right]^{-z\nu} \qquad (6).$$

Our data can be fitted (figure 3) with plausible critical exponent values, $z\nu = 8.5 \pm 0.3$ and $\tau_o^* = 1 \pm 0.5$ μsec in the aligned ferrofluid and $z\nu \approx 7.5 \pm 0.3$ and $\tau_o^* \approx 1 \pm 0.5$ μsec in the random one. The large value of $\tau_o^*$ (~1 μsec) can be easily explained in terms of the Arrhenius-Néel law: $\tau_o^*(T) \sim \tau_o exp\{E_a/k_B T\}$. With $\tau_o \sim 10^{-9}$ s and $E_a/k_B = 2 \times 300$ K, $\tau_o^*$ at $T_g = 70$ K reaches the order of microseconds. Thus, it appears that the superspin glass transition is not lost by the anisotropy-axis alignment of the ferrofluid but with the critical exponent that is slightly higher than its randomly oriented counterpart. Also, unlike the glass transition determined from static susceptibility, $T_g(\omega \neq 0)$ values are found to behave differently in the aligned and the randomly oriented states. It may be worth noting that in atomic spin glasses, the observed critical exponent ($z\nu$) is larger in Ising spin glasses than in Heisenberg-like spin glasses [9].

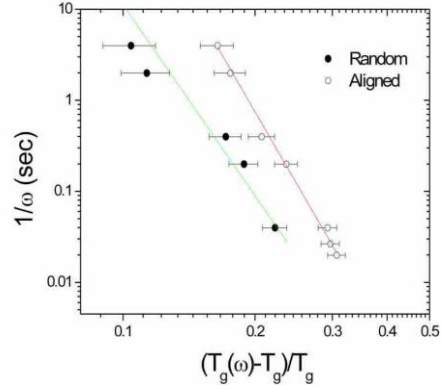

**Figure 3.** Displacement of transition temperature with frequency determined from in-phase ac susceptibility in a ferrofluid with and without anisotropy-axis alignment. The critical exponent, appearing as the slope on the log-log scale is slighter larger in the aligned ferrofluid.

*3.3. SSG Aging in the very low field limit*

We now discuss the effect of anisotropy axis alignment on the aging behaviour in the low temperature superspin glass states. Let us start by comparing the relaxation rate distribution spectra, $S(t) = dM/d\log(t)$ between the two systems. Examples of $S$ spectra taken at 0.7 $T_g$ with $t_w$= 3 ks in both systems are presented in figure 4 (top panel). As can be seen from the graph, the peak ($S_{max}$) width of relaxation rate in the aligned SSG state is considerably narrower than that in the random SSG state. This may not come as a surprise considering that the anisotropy energy distribution of a uni-axial, single-domain nanoparticle system depends on the distribution of angles between the constituting particles' magnetization and the external field directions. Thus, the distribution of energy barriers of correlated superspin domains should be concentrated about a common value in the aligned SSG state.

In Figure 4 (bottom panel), the $S_{max}(t)$ locations, $t_w^{eff}$, obtained from the ZFCM relaxation rate curves are plotted against the experimental waiting time, $t_w$, on a *log-log* scale for both SSG states. These measurements were performed at $T_m = 49$K (~$0.7T_g$) with the excitation field $H = 0.5$ Oe and $t_w$ was varied between 3 and 24 ks. As discussed in section 3.3, in the low field limit, one expects to obtain $\log(t_w^{eff}) = \log(t_w)$. As can be seen from the figure, $t_w^{eff}$ is $\approx t_w$ in the random SSG state. On the other hand, the values of $t_w^{eff}$ of the aligned SSG state are larger than the experimental $t_w$ by approximately 1500 sec. By adding an extra time, $t_{ini}$, to $t_w$; $t_w \to t_w + t_{ini}$, with $t_{ini} \approx 1500$ s, the $t_w^{eff}$ plot of the aligned SSG state coincides with that of the random state. The presence of $t_{ini}$ may indicate that the *aging* had started during the cooling, *i.e.*, ~1500 s prior to the experimentally defined quench time, but only in the aligned SSG state despite the identical cooling rate used in both experiments.

Similarly in atomic spin glasses, an enhanced sensitivity to cooling rates, also known as a 'cumulative aging' effect; that is, a tendency for aging to pile up from one temperature to another, have been observed in Ising systems [42,43]. The effective age of an Ising spin glass *increased* after slower cooling, while Heisenberg spin glasses remained nearly insensitive to the same cooling-rate variations [9]. This analogy is particularly appealing as the anisotropy axis alignment should qualitatively drive the system toward an Ising-like magnetic state. Is is also consistent with the critical exponent analysis in the previous section where the critical exponent, $z\nu$, associated with the aligned SSG transition was found to be larger than in the random case.

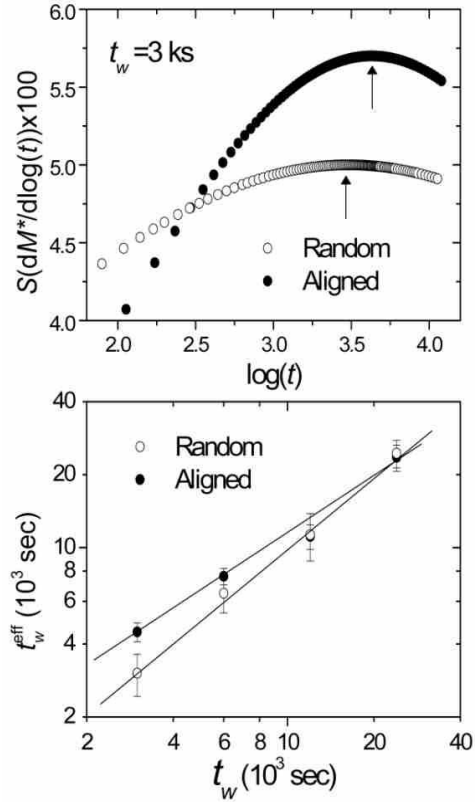

**Figure 4**. (Top) Relaxation rate of ZFCM, $S$, vs. log($t$) in anisotropy-axis aligned and random SSG states with an external field of 0.5 Oe and the waiting time ($t_w$) of 3 ks. The arrows indicate the positions of $S_{max}$. (Bottom) $t_w^{eff}$ vs. $t_w$ found in the ZFCM relaxation curves at 0.5 Oe and with $t_w = 3, 6, 12$ and 24 ks on a log-log scale.

*3.4. Magnetization scaling*

Next, we examine the ZFCM scaling of aligned and randomly oriented ferrofluids with $t_w$ values ranging from 3 to 24ks and under 0.5Oe. As mentioned above, the subtraction of the superparamagnetic ($m_{SPM}$) and the equilibrium ($m_{eq}$) components is necessary in order to achieve a good scaling [6,30]. These contributions follow the forms; $B(\log(t/\tau_o^*))$ and $-A(t/\tau_o^*)^{-\alpha}$, respectively, where $B$ and $A$ are prefactors and $\alpha$ is a scaling exponent. The value of $\tau_o^*$ is fixed according to the Arrhenius-Néel law as described in section 3.2. The corresponding $\tau_o^*$ values at 49 and 59.5 K are, 200 and 26μs, respectively. The fitting parameters used to scale the ZFCM curves are summarized in table 1 and the corresponding scaling curves are shown in figure 5.

The most remarkable difference between the two scaling curves at 49 K is the critical exponent "$\mu$" in the scaling variable $\lambda/t_w^{\mu}$ (see section 1). $\mu = 0.91$ found in the random SSG is close to the values found in atomic spin glasses [28] as well as the results obtained in more concentrated maghemite ferrofluids [6]. On the other hand, in the aligned SSG state $\mu$ has been shifted to a dramatically smaller value, 0.61. In atomic spin glasses; if $\mu = 1$ ($t_w^{eff} = t_w$) then the system is termed *fully aging*, if $\mu = 0$ then there is no aging (*i.e.,* magnetization relaxation does not depend on $t_w$) and in-between values of $\mu$ reflect 'subaging' [44,45]. Therefore, the $\mu$ value close to unity found in the randomly oriented SSG confirms the earlier observation $t_w^{eff} \propto t_w$. The results also agrees with the smaller slope found in Figure 4 (bottom panel) for the aligned SSG state and it may also reflect, partly, the cooling rate effect as discussed above.

**Table 1.** Fitting parameters used for the ZFCM scaling. Note that due to a multiple number of fitting parameters, slightly different solutions to $A$, $B$ and $\alpha$ can equally produce reasonable scaling. However, $\mu$ is the most influential on the overall scaling quality and it must be close to the values indicated below.

|   | Random 49K | Aligned 49K | Aligned 59.5K |
|---|---|---|---|
| $A$ | 0.26 | 0.26 | 0.25 |
| $\alpha$ | 0.22 | 0.07 | 0.09 |
| $B$ | 0.001 | 0.005 | 0.015 |
| $\mu$ | 0.91 | 0.61 | 0.29 |
| $\tau_o^*$ | 200 μs | 200 μs | 26 μs |

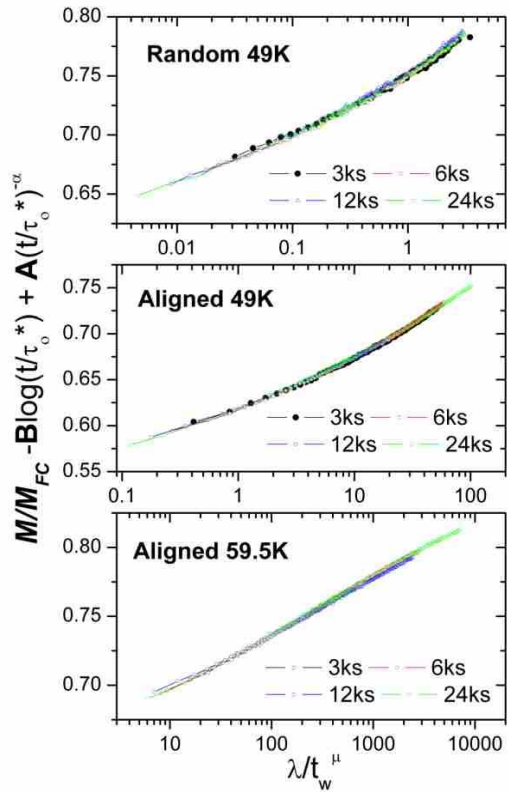

**Figure 5.** Scaling of ZFCM relaxation curves obtained at 49 K in random (top) and aligned (middle) SSG states and at 59.5 K in aligned SSG state (bottom) with $t_w = 3\sim24$ ks. A superparamagnetic contribution [$B\log(t/\tau_o^*)$] and an equilibrium contribution [$-A(t/\tau_o^*)^{-\alpha}$] are subtracted from the total ZFCM. See text for details.

We have also attempted to scale the ZFCM data obtained at 59.5K (0.84 $T_g$) in the aligned SSG phase (figure 5, bottom panel). Due to the higher temperature toward $T_g$, a larger proportion of the total magnetization grew within the first few seconds immediately following the external field application, before we could perform our first measurement with our current experimental set-up. Consequently, the range of magnetization change became much smaller than those probed during the measurements at 49 K. Nevertheless, we were still able to achieve scaling using the same data treatment but with two marked differences. First, the $B$-term corresponding to the contribution from time-logarithmic superparamagnetic particles grew larger; $B(59.5K) \sim 0.015$ as opposed to $B(49K) \sim 0.005$. Second, the scaling exponent $\mu$ is further reduced to 0.29! In Heisenberg spin glasses, the value of $\mu(T)$ has a plateau like structure around $\mu \sim 0.9$ across a wide range of temperature between 0.5 and 0.9 $T_g$. $\mu(T)$

then falls off rapidly as the system approaches the critical region near the glass transition temperature; $T > 0.9\, T_g$ [46]. In an Ising spin glass, the cumulative aging effect, which pushes $\mu$ towards smaller values in isothermal aging experiments, was tentatively attributed to its more extended critical region compared to conventional Heisenberg spin glasses [42]. A similar phenomenology akin to the cumulative aging is perhaps present in an aligned frozen ferrofluid system. Additional magnetization relaxation measurements (ZFCM or TRM) are needed to test if the $\mu(T)$ drop-off occurs at a lower temperature (in $T_g$) in a frozen ferrofluid superspin glass phase.

*3.5 Zeeman Energy*

Lastly, we focus our attention on the effective age ($t_w^{eff}$) change due to the application of $H$; that is, the Zeeman energy coupled to dynamically correlated superspins. In figure 6, the effective times, $t_w^{eff}$ measured at different $t_w$ values are plotted as functions of magnetic field. As $\ln(t_w^{eff}) \sim E_Z/k_BT$, a semi-log plot of $t_w^{eff}$ vs. $H$ depicts equivalently the Zeeman energy dependence on $H$. The difference in the $t_w^{eff}$ dependence on $H$ between the two SSG states is very clear. For a randomly oriented ferrofluid, we confirm our previous observation that $t_w^{eff}$ shows a near quadratic field dependence. In a stark contrast to this, $t_w^{eff}$ of an aligned ferrofluid shows a close-to-linear dependence. Even at the 59.5 K where the relaxation was found to be much faster than at 49 K, the linear dependence of $t_w^{eff}$ is still clear (see figure 7). Once again, the Zeeman energy dependence of $H$ in a random and an aligned SSG states resembles that of Heisenberg ($H^2$) and Ising ($H$) spin glasses, respectively [11,12].

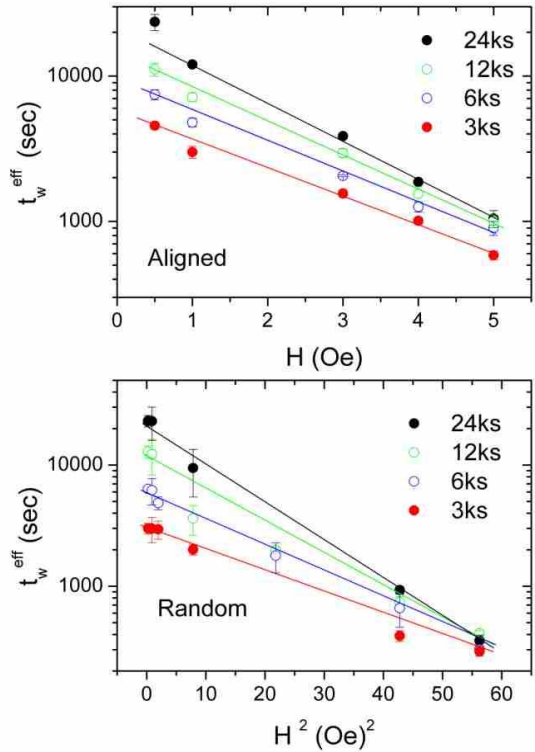

**Figure 6:** Effective age of the sample dependence on external magnetic field at 49 K. $t_w^{eff}$ was found to depend linearly in the aligned SSG state (top) while in the random SSG state, it exhibited near $H^2$ dependence.

In order to extract the typical number of dynamically correlated spins, $N_s(t_w)$, a more careful examination on the forms of $E_Z$ and their interpretations is required. For example, although the $t_w^{eff}$ vs $H$ curves of the random SSG state on the log-log scale show near $H^2$ dependence, it is not purely so. In Heisenberg spin-glasses, the quadratic dependence of $E_z$ has been phenomenologically associated to $N_s\chi_{FC}H^2$. While this interpretation may very well be valid in atomic spin glasses whose field range of investigation exceeds 1000G [11], it may not be adequate for a superspin glass because the low field range (where the ZFCM approach is valid) is limited to $H < 10$ G due to a large magnetic moment of nanoparticles. The effective local field due to dipolar interactions, *e.g.,* from near-by large nanoparticles that are too large to relax within a laboratory time scale, may significantly alter the $N_s$ value to be determined. Furthermore, the possibility of another entirely different aging mechanism specific to slowly interacting dipolar fine magnetic particles should also be considered [21,22]. These analysis are currently underway to extract realistic $N_s$ values.

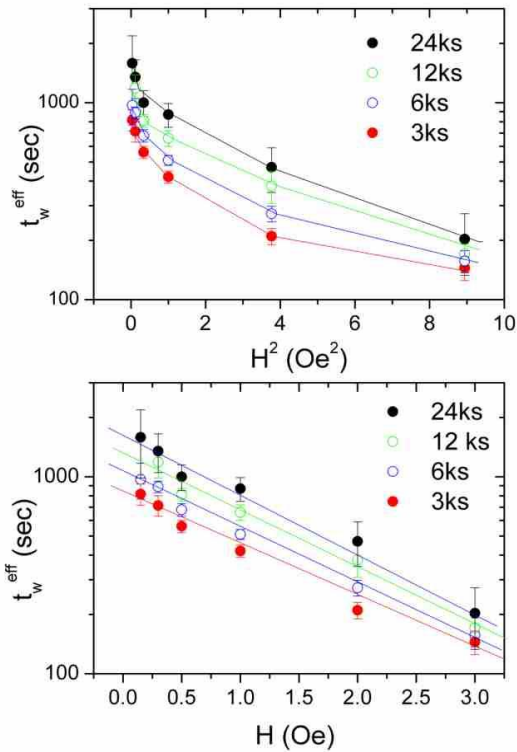

**Figure 7**. Effective age of the aligned sample vs. $H^2$ and $H$ at 59.5 K. Linear relationship between $t_w^{eff}$ and $H$ is clearly observed.

IV. CONCLUSION

We have investigated the effect of the magnetic anisotropy-axis alignment in the superparamagnetic and the superspin glass states of a frozen ferrofluid. The anisotropy axis alignment was achieved by means of strong (> 15 kOe) magnetic field applied to a ferrofluid in its liquid state. In the high temperature SPM state, the linear susceptibility of aligned ferrofluid increased by a factor of 2 ~ 4 with respect to that measured in the randomly oriented state. The SSG transition temperature extracted from the linear magnetic susceptibility curves, $\chi(T)$, remained insensitive to the anisotropy axis alignment. Additionally, $\chi(T)$ fit to the Curie-Weiss law in the high temperature SPM regime revealed the negative interaction energy to be similar in both states.

The low temperature superspin glass dynamics explored via ac susceptibility and 'zero-field cooled magnetization' (ZFCM) relaxation measurements, however, show distinct differences in the out-of-equilibrium dynamics of SSG phase due to the anisotropy-axis alignment. These changes are:

a) Larger critical exponent in an aligned ferrofluid. $T_g(\omega)$ was also found to be larger in the aligned system for all $\omega$ values explored.
b) Subaging-like behaviour in the aligned SSG state. The effect appeared only in the aligned sample as an initial age and as a smaller scaling exponent, μ (~0.9 in the random SSG state to ~ 0.6 in the aligned SSG state at 0.7 $T_g$).
c) Zeeman energy dependence on $H$. $E_Z$ depends linearly in the aligned SSG state, while near-quadratic dependence was observed in the random SSG state.

Interestingly many of these above listed differences between the anisotropy-axis aligned and the randomly oriented SSG states resemble those found in Ising-like and Heisbenberg spin glasses.